\RequirePackage{fix-cm}
\documentclass[smallextended]{svjour3}       
\smartqed  
\usepackage{graphicx}
\usepackage{tabularx}
\usepackage{booktabs}
\usepackage{microtype}
\usepackage{graphicx}
\usepackage{enumitem}

\usepackage{hyperref}
\usepackage{float}
\usepackage{titlesec}
\usepackage{caption}
\usepackage{subcaption}

\journalname{Journal of Signal Processing Systems}
\begin{document}

\title{Design, Implementation, Comparison, and Performance analysis between Analog Butterworth and Chebyshev-I Low Pass Filter Using Approximation, Python and Proteus
}


\author{Navid Fazle Rabbi
}


\institute{Navid Fazle Rabbi \at
              Dhaka, Bangladesh \\
              Tel.: +880-1706319096\\
              \email{navidfazlerabbi@iut-dhaka.edu}           
}

\date{Received: 24 January, 2021 / Accepted: date}

\maketitle

\begin{abstract}
Filters are broadly used in signal processing and communication systems in noise reduction. Butterworth, Chebyshev-I Analog Low Pass Filters are developed and implemented in this paper. The filters are manually calculated using approximations and verified using Python Programming Language. Filters are also simulated in Proteus 8 Professional and implemented in the Hardware Lab using the necessary components. This paper also denotes the comparison and performance analysis of filters using Manual Computations, Hardware, and Software. 
\end{abstract}

\section{Introduction}
\label{intro}
Filters play a crucial role in the field of digital and analog signal processing and communication networks. The standard analog filter architecture consists of two main components: the problem of approximation and the problem of synthesis. The primary purpose is to limit the signal to the specified frequency band or channel, or to model the input-output relationship of a specific device.\cite{singh2011determining}\cite{malica2011design}

This paper revolves around the following two analog filters:
\begin{itemize}
    \item Butterworth Low Pass Filter
    \item Chebyshev-I Low Pass Filter
\end{itemize}

These filters play an essential role in alleviating unwanted signal parts. Each of them has their use-case and does well in such situations.

\section{Filters}

\subsection{Butterworth Filter}
The Butterworth filter has a maximum flat response, that is, no passband ripple and a sharp roll-off. It is known as the \textbf{"flat maximum magnitude"} filter.\cite{katiyar2012design} The increase in order results in a more non-linear phase response of the Butterworth Filter. Two parameters mathematically define this filter; the cut-off frequency and the number of poles. The frequency response is at most smooth in the passband and rolls to zero in the stopband.\cite{babu_2007}

The magnitude squared response of low pass Butterworth filter-

\begin{equation}
    H(j\omega) = \frac{1}{1 + (\frac{\omega}{\omega_{c}})^{2n}}
\end{equation}

Butterworth filters are used as high-frequency Low Pass filter detection, smoothing filters, and anti-aliasing filters.

\subsection{Chebyshev-I Filter}
Chebyshev filters may be analog or digital filters; Chebyshev comes in two types: Chebyshev Type-I and the other is called Chebyshev Type-II. Chebyshev Type-I filter has a steeper roll and more passband ripple. Chebyshev minimizes the error between the ideal and the real filter characteristics. As the ripples are present in the passband, Chebyshev has less ripple response in the passband and a more erratic response in the stopband. The response of the Chebyshev filter process is weak.\cite{proakis2004digital}

The Chebyshev Type reduces the absolute difference between the ideal and the practical frequency response over the entire passband. It filters by adding the same ripple in the passband. The transition is steeper from the passband to the stopband comparing to the Butterworth filter.\cite{sandhu2016study}

The magnitude squared response of low pass Chebyshev-I filter-

\begin{equation}
    H(j\omega) = \frac{1}{1 + \varepsilon^{2}C_{n}^{2}(\frac{\omega}{\omega_{p}})}
\end{equation}

\section{Softwares and Programming Language}

\subsection{Proteus 8 Professional}
The Proteus Design Suite is used to design the filters and simulate to get the results.\cite{proteus}

\subsection{Python}
Python is a high-level, and general programming language. Python is used to decide if our measurements are accurate and to map the appropriate readings on the graph. It has been used to compare the filters as well.\cite{van1995python}\cite{kuhlman2009python} Python Packages used: \textbf{SciPy}\cite{2020SciPy-NMeth}, \textbf{NumPy}\cite{2020NumPy-Array}\cite{harris2020array}, \textbf{Matplotlib}\cite{hunter2007matplotlib}.

\section{Design and Implementation of Analog Butterworth Low Pass Filter}

\subsection{Approximation}

\cite{nair2004digital}When $n$ stages are cascaded, we get the $n$th order filter. The higher the value of $n$(order), the sharper cut-off. For $n$th order filter (Butterworth Polynomial of Order $n$) -

\begin{equation}
    \left |H(\omega)  \right | = \left [\frac{1}{1+ (\frac{\omega}{\omega_{c}})^{2n}}  \right ]^{\frac{1}{2}}
\end{equation}

If $H\omega$ is expressed in decibels-

\begin{equation}
    \left |H(\omega)  \right | in\:dB = -10log\left [1+(\frac{\omega}{\omega_{c}})^{2n}  \right ]
\end{equation}

This is the design equation, which has been used throughout. 

\subsubsection{Design Introduction}
Specifications that we used to efficiently design an Analog Low Pass Butterworth Filter-
\begin{itemize}
    \item A passband gain of $-0.5\:dB$ was used.
    \item $100\:rad/s$ was assumed for the frequency up to which the pass band gain remained more or less constant.
    \item An attenuation of $-20\:dB$ was used.
    \item From $200\:rad/s$ the attenuation was started.
\end{itemize}

\subsubsection{Design Procedure}
\begin{itemize}
    \item \textbf{Calculating the Values of $n$ and $\omega_{c}$:} Using the given values of $H_{1}(\omega)$ and $H_{2}(\omega)$, we can solve from the equation $(3)$ and get the desired values: $n$ and $\omega_{c}$. If the values are in decibel, we can use the equation $(4)$.
    \item \textbf{Determining the Poles of $H(s):$} Low Pass Butterworth filters are all-pole filters. We use the following equations to calculate the number of poles and the angles between them: Number of Butterworth Poles $=\:2n$, Angle between the Poles, $\theta = \frac{360^{\circ}}{2n}$. If the order of the filter, $n$ is an even number, the location of the first Pole will be $\frac{\theta}{2}$, with respect to the $x$-axis. Whereas, if the order of the filter, $n$ is an odd number, the location of the first Pole will be on the $x$-axis.
   \item \textbf{Determining the Valid Poles of $H(s)$:} The left half of the S-Plane are alone stable poles. Whereas, the right half poles are unstable. With respect to the $x$-axis, Poles between $90^{\circ}$ and $270^{\circ}$ alone are valid poles. And if $\alpha$ is the angle of a valid pole, then the pole and its conjugate are located, with respect to $x$-axis at
        \begin{equation}
            \left [\omega_{c}(cos\alpha \pm jsin\alpha)  \right ]
        \end{equation}

    \item \textbf{Finding the Expression of $H(s)$:} Transfer Function can be generated from the calculations obtained.
    In case of a first order filter, the transfer function will be-
    \begin{equation}
        H(s) = \frac{\omega_{c}}{s + \omega_{c}}
    \end{equation}
    In case of second order filter, the transfer function will be of this format-
    \begin{equation}
        H(s) = \frac{\omega_{c}^{2}}{(s+a+jb)(s+a-jb)}
    \end{equation}
    $a + jb$ and $a-jb$ being the poles, determined earlier.
    \item \textbf{Determination of the Circuit Components:} The filter structure that is used is the Sallen-Key Structure. Sallen-Key filters are second-order RC filters containing op-amps.\cite{sallen6circuit}\cite{razavi2013fundamentals}
\end{itemize}

\subsection{Manual Mathematical Calculations}

\begin{itemize}
    \item \textbf{$n$ and $\omega_{c}$:} With the specified values taken, the order of the filter, $(n)$ and the cut-off frequency $(\omega_{c})$ is calculated with the equation $(4)$. The results are, $n = 4.967 \approx 5$ and $\omega_{c} = 123.41$.
    \item \textbf{Valid Poles:} Calculating with the obtained values, we get, the Number of Poles = $10$ and the angle between each of the pole $= 36^{\circ}$. But not all of them are valid poles. Also, the order of the filter is odd, which means the location of the first pole is on the $x$-axis. We know that the angles that lie between $90^{\circ}$ and $270^{\circ}$ are valid poles. So, using the equation $(5)$, the valid poles are: \textbf{$-38.134 \pm j117.363$, $-99.84 \pm j72.565$, $-123.41$.}
    \item \textbf{Transfer Function $H(s)$:} From the calculations done, the transfer function stands,
    \begin{equation}
        \resizebox{.78\hsize}{!}{
        $H(s) = \frac{123.41 \times 123.41^{2} \times 123.41^{2}}{(s + 123.41)(s^{2}-76.268s + 15345.887)(s^{2}-199.68s + 15233.705)}$}
    \end{equation}
    \item \textbf{Circuit Components:} Comparing equation $(8)$, with the equation $(6)$ for the First-order filter components and with the equation $(7)$ for the Second-order filter components, we get, \textbf{1st Second-Order Filter:} $R=10k\Omega$, $C_{1}=212.42nF$, $C_{2}=20.135nF$, \textbf{2nd Second-Order Filter:} $R=10k\Omega$, $C_{1}=81.168nF$, $C_{2}=53.13nF$, \textbf{First-Order Filter:} $C=0.1\mu F$, $R=81.03k\Omega$.

\end{itemize}

\subsection{Circuit Design}
The computed components were assembled to create a Fifth-Order Butterworth Analog Low Pass filter. The following circuit was constructed with components and tested in the Hardware Lab and with software simulations.

\begin{figure*}[h!]
    \centering
    \includegraphics[width=.8\textwidth]{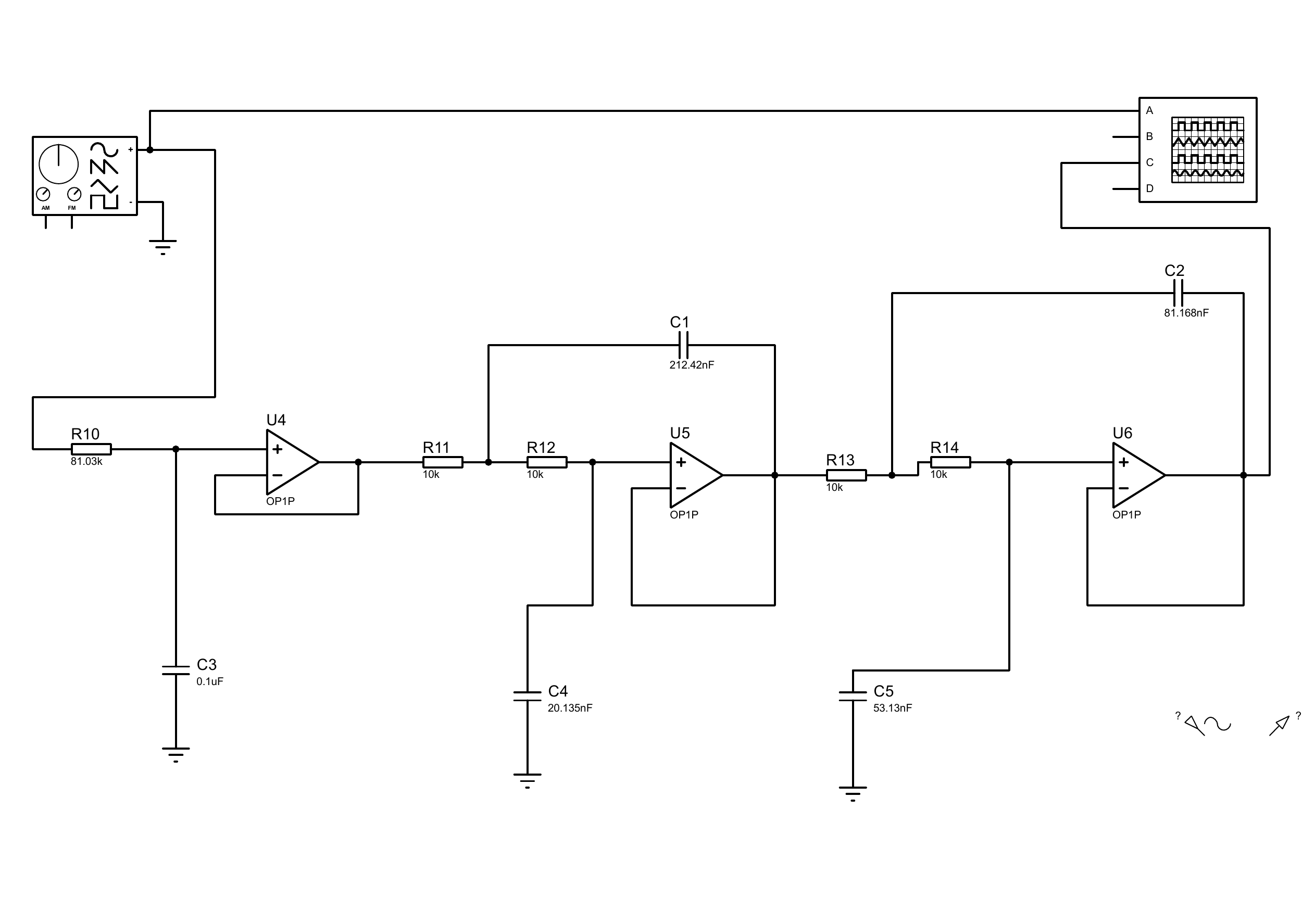}
    \caption{Analog Low Pass Butterworth Filter designed with the computed components}
    \label{Butterworth Filter Circuit}
\end{figure*}

\subsection{Software Implementation}
\subsubsection{Proteus Implementation}
The circuit is constructed in the proteus environment, using the mathematically computed filter components. The circuit is then connected to the input voltage signal generator, where a square signal is fed. The value of input frequency was 15.91Hz $\approx$ 100 rad/s. The output voltage of the circuit is observed using a digital oscilloscope, which gives a desired result. The Frequency Response shows desired output by the simulation as well.
\begin{figure}[H]
    \centering
    \subfloat[\centering Circuit Simulation in Proteus 8]{{ \includegraphics[width=.5\textwidth]{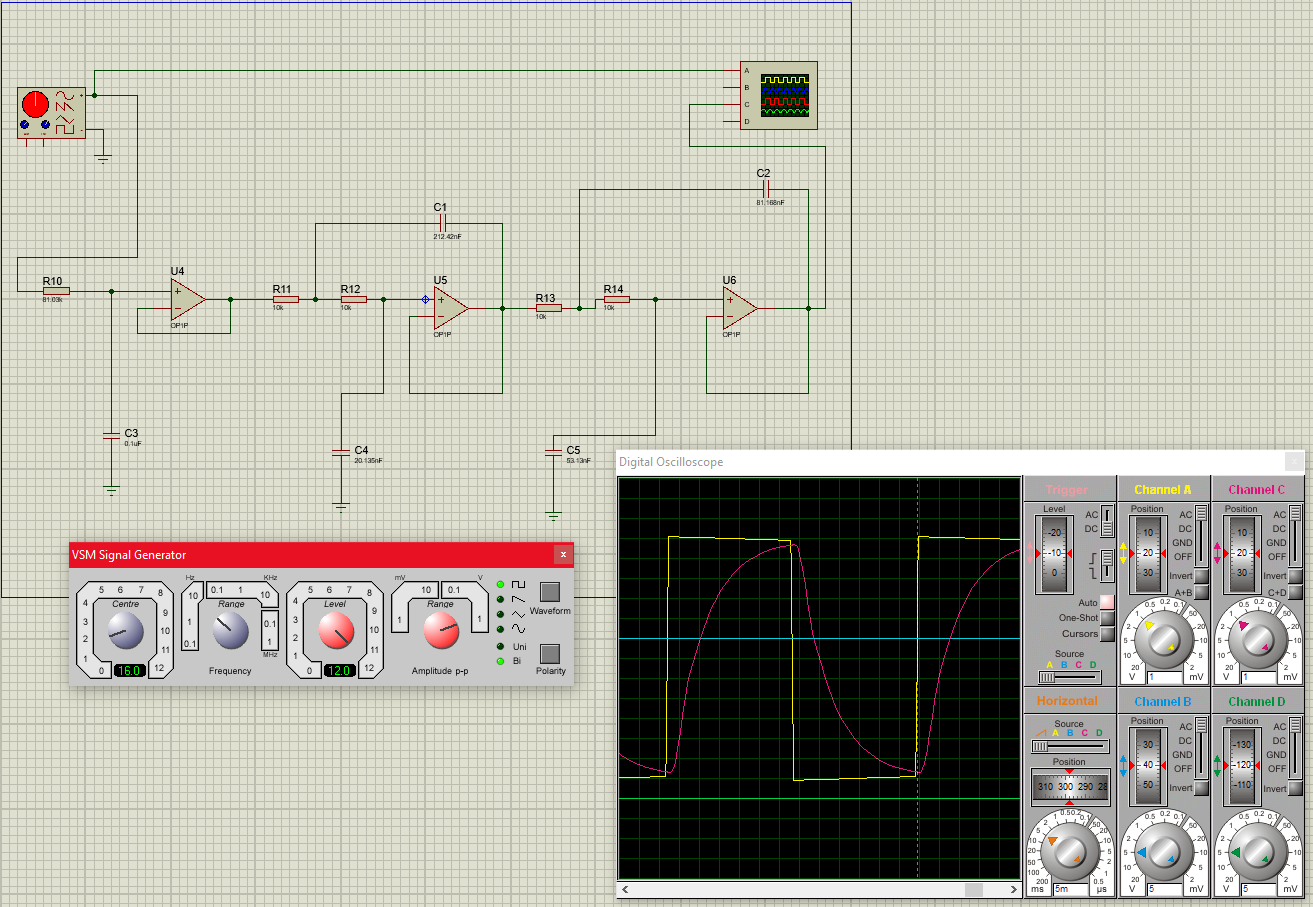} }}%
    \qquad
    \subfloat[\centering Output at the Output Voltage End in Proteus 8]{{ \includegraphics[width=.3\textwidth]{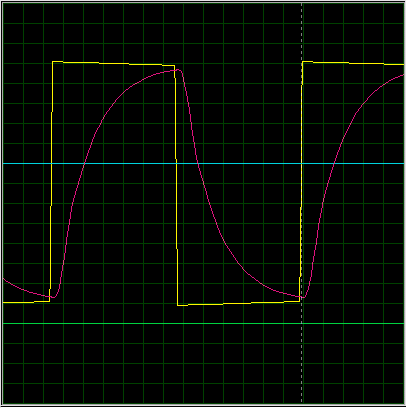} }}%
    \qquad
    \subfloat[\centering Simulated Frequency Response of the Designed Butterworth Filter]{{\includegraphics[width=.5\textwidth]{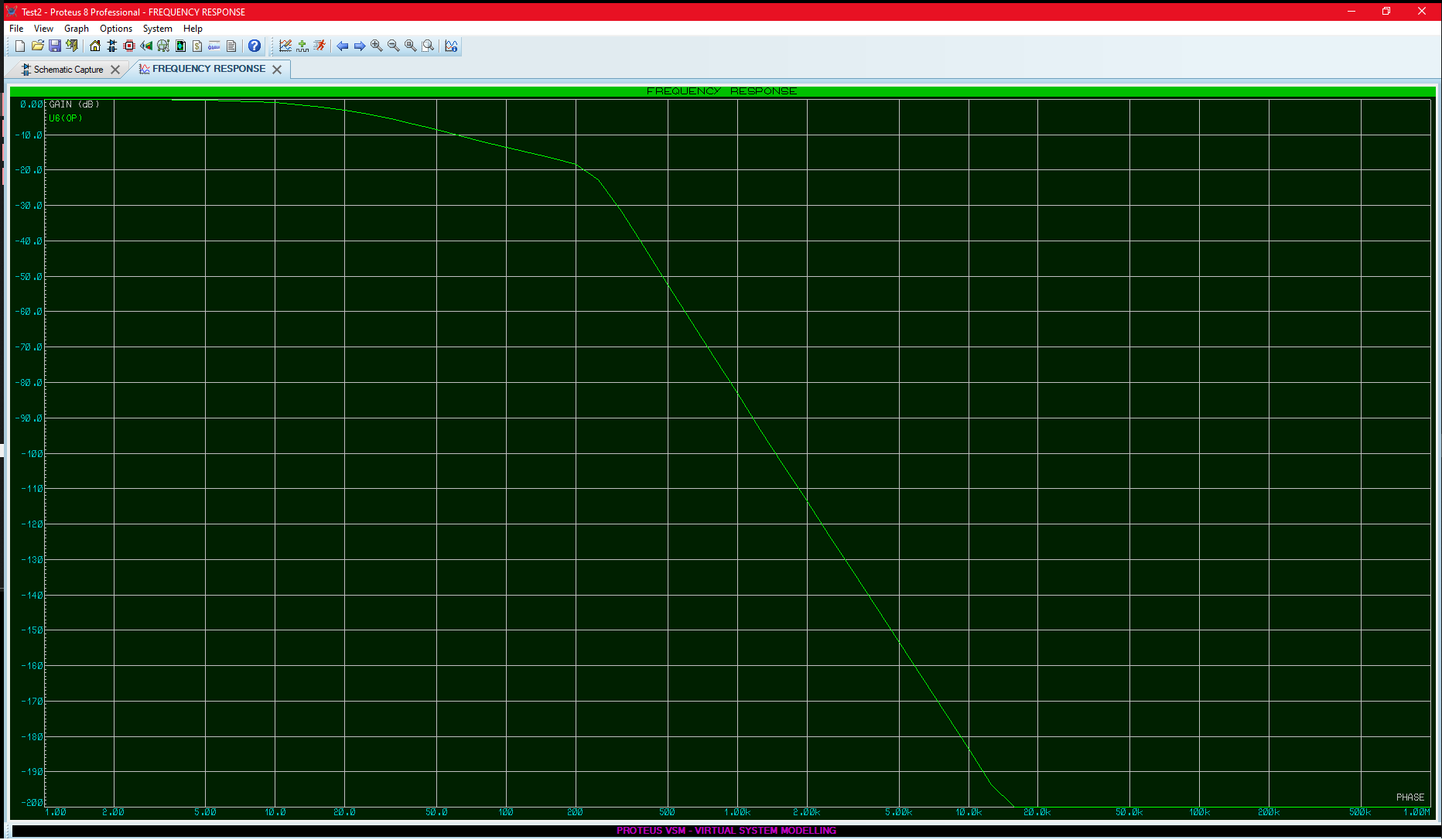} }}%
    \caption{Software Outputs}%
    \label{Butterworth}%
\end{figure}

We can conclude by saying that our circuit simulation produced the desired output.

\subsubsection{Python}
\begin{itemize}
    \item \textbf{Order and Cut-off Frequency Determination:}
    
    \textbf{scipy.signal.buttord} function, gives us, \textbf{Order of the Filter}, $N:\:5$, and \textbf{Cut-off Frequency}, $Wn:\:123.4120164$.

    The output matches the results that were calculated earlier.
    
    \item \textbf{Pole-Zero Location Determination:}
    \textbf{scipy.signal.butter} function has been used to get the pole, zero locations of the Butterworth filter. Using this function we get, \textbf{Pole Locations:} $-38.13641037 \pm 117.37180235j$, $-99.84241855 \pm 72.53976317j$, $-123.41201636$. 
    
    \item \textbf{Pole and Zero Location Mapping:} \textbf{plotPoleZeros} function is created to map the pole-zero location on an S-plane.
 
    The results show the stable pole locations on the left of the imaginary axis, which matches our calculations distinctively. As the order of the filter is odd, there will be pole location on the $x$-axis. The map shows a pole location on the $x$-axis. So, the mapping is correct.
    
    \item \textbf{Frequency Response Mapping:} \textbf{scipy.signal.freqs\_zpk} function has been used to get the frequency response of the Butterworth filter designed, which shows accurate results.
    \end{itemize}
    
    \begin{figure}[H]
    \centering
    \subfloat[\centering Pole Zero Location mapping with \textbf{plotPoleZeros} function]{{ \includegraphics[width=.4\columnwidth]{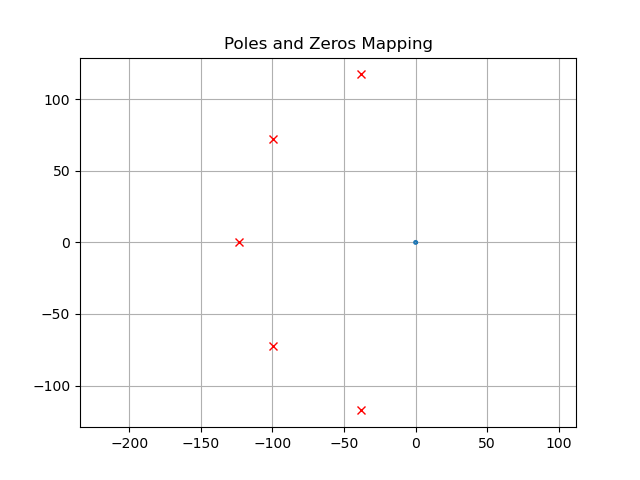} }}%
    \qquad
    \subfloat[\centering Frequency Response of the Butterworth Filter]{{\includegraphics[width=.4\columnwidth]{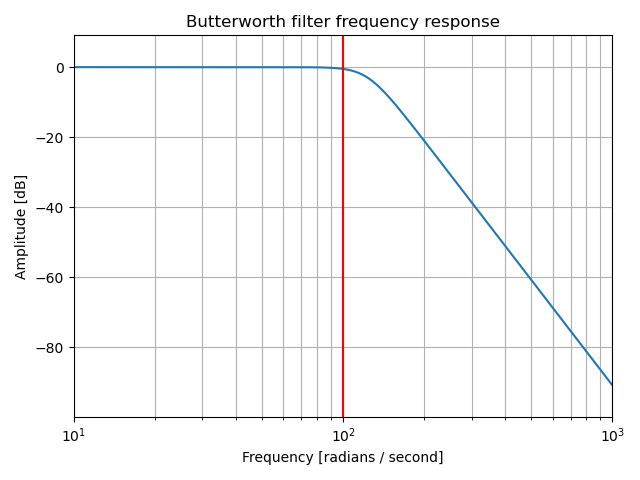} }}%
    \caption{Python Outputs}%
    \label{Butterworth}%
\end{figure}

\subsection{Hardware Implementation}
The designed circuit was implemented with hardware components in the lab, which gave the following voltage output results.

\begin{figure}[H]
    \centering
    \subfloat[\centering Output at input frequency of 9 Hz]{{ \includegraphics[width=.4\columnwidth]{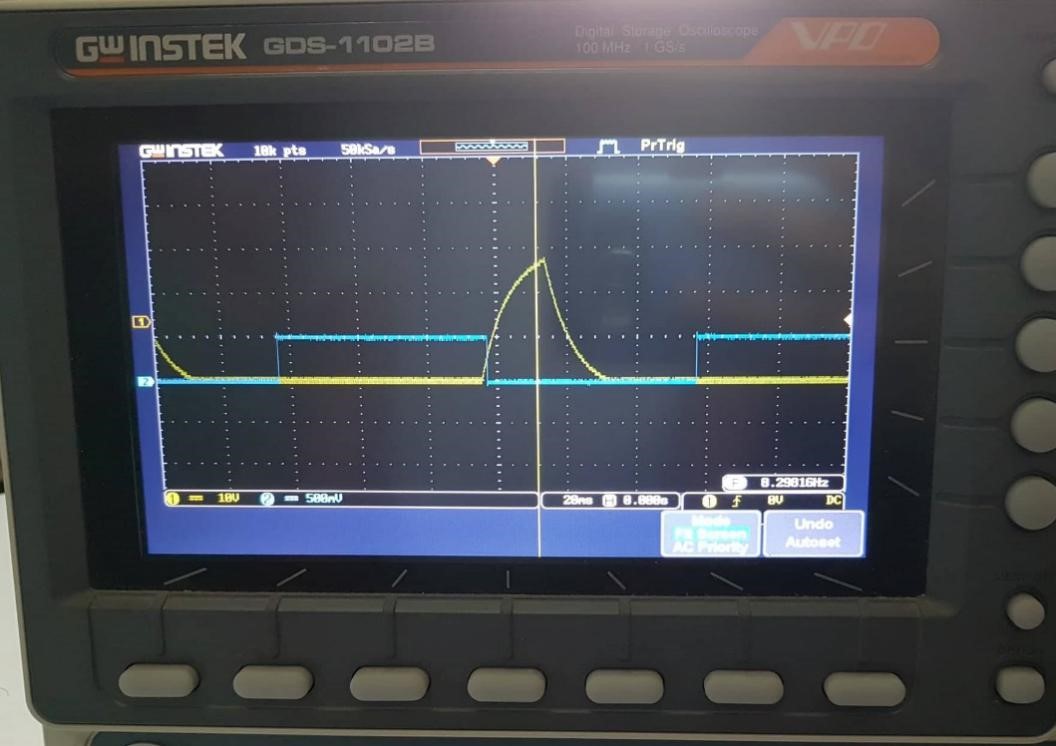} }}%
    \qquad
    \subfloat[\centering Output at input frequency of 15 Hz]{{\includegraphics[width=.4\columnwidth]{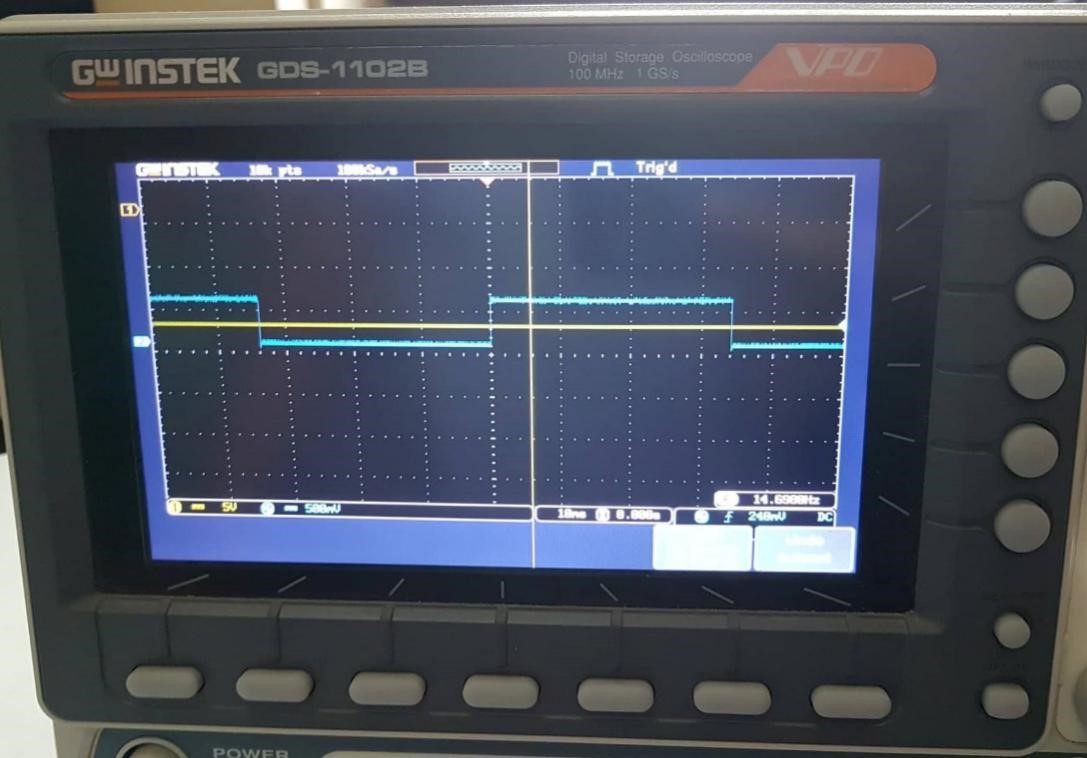} }}%
    \caption{Hardware Outputs (Blue = input, Yellow = output)}%
    \label{Butterworth}%
\end{figure}

\section{Design and Implementation of Analog Chebyshev-I Low Pass Filter}

\subsection{Approximation}
\cite{nair2004digital}Chebyshev approximation is used when passband gain contains ripples. The Chebyshev Polynomial-

\begin{equation}
    \left | H(\omega) \right |^{2}=\frac{1}{1+\varepsilon^{2}C_{n}^{2}}
\end{equation}

where, $\varepsilon =$ amount of ripple in the magnitude, and $C_{n} =$ Chebyshev coefficient, given by

\begin{equation}
    C_{n}=cosh(n\:cosh^{-1}\frac{\omega}{\omega_{c}})
\end{equation}
The Chebyshev Polynomials can be expressed in decibels as well.

\begin{equation}
    \left | H(\omega) \right |\:in\:dB =-10log(1+\varepsilon^2C_{n}^{2})
\end{equation}

\subsubsection{Design Introduction}
Same specifications that we used to design an Analog Low Pass Butterworth Filter was used to design the Chebyshev-I Filter. In case of Chebyshev design, $\omega_{1}=\omega_{c}$.

\subsubsection{Design Procedure}

\begin{itemize}
    \item \textbf{Calculating the Value of $n$:} Given the equation $(9)$, at $\omega_{1} = \omega_{c}$, the value of $C_{n}$ is $1$. For the value of $H_{1}(\omega)$, we can compute the value of $\varepsilon$ from this equation. Similarly, using the value of $\varepsilon$ obtained, and with the $H_{2}(\omega)$, we can calculate the value of $C_{n}$ from this equation as well. Then with the value of $\varepsilon$ and $C_{n}$, we can get the order of the filter, $n$ using the equation $(10)$.
    \item \textbf{Determining the Poles of $H(s)$:} First we need to calculate the valid Butterworth poles, with obtained order of the filter, $n$. By Butterworth poles, it is meant that, we should calculate the poles of a Butterworth filter with this value.\cite{parks1987digital} Then, we need to calculate the factor $k$, from the following expression
        
        \begin{equation}
            k = \frac{1}{n}\:sinh^{-1}(\frac{1}{\varepsilon})
        \end{equation}
        
     After obtaining $k$, we need to find the values of $tanh\;k$ and $cosh\;k$.Then, we need to multiply the real part of the butterworth poles with $tanh\;k$. This is known as normalization. Now, the poles are denormalized by multiplying with $cosh\;k$ to get the desired Chebyshev poles.
    
    \item \textbf{Finding the Expression of $H(s)$:} After obtaining the poles, we can easily get the transfer function of the Chebyshev-I filter. The transfer function-
    
        \begin{equation}
             H(s) = \frac{P}{(s+C_{1})(s+C_{2})(s+C_{3})(s+C_{4})}
        \end{equation}
    
    where, $P$ is the magnitude of the poles, which is given by- 
    
    $P = \left \{ \left | C_{1} \right |\left | C_{4} \right | \right \}\left \{ \left | C_{2} \right |\left | C_{3} \right | \right \}$. And $C_{1}, C_{2}, C_{3}, C_{4}$ are the Chebyshev Poles. 
    
    \item \textbf{Determination of the Circuit Components:} The procedures to calculate the circuit components, are same as the Butterworth circuit component calculation. 
\end{itemize}

\subsection{Manual Mathematical Calculations}
\begin{itemize}
    \item \textbf{$n$:} With the specified values taken, the order of the filter, $(n)$ is calculated. First the value of $\varepsilon$ is calculated, and then it is used to obtain the value of $C_{n}$. Then from the equation $(10)$ we get the value of $n = 3.07 \approx 4$.\cite{nair2004digital}.
    \item \textbf{Valid Poles:} The valid butterworth poles are calculated with the obtained order of the filter. Then, from equation $(12)$ we get $k=0.44$. With this value of $k$, we get, $tanh\;0.44=0.417$ and $cosh\;0.44=1.1$. Normalizing the Butterworth poles and then again denormalizing it, gives us the Chebyshev Poles: $-17.567 \pm j101.64$, $-42.383 \pm 42.13$
    \item \textbf{Transfer Function $H(s)$:} Now using the equation $(13)$, we can obtain the Transfer Function of the Cheybeshev Filter according to the specifications,
    
    \begin{equation}
    \resizebox{.78\hsize}{!}{
    $H(s) = \frac{37995628.25}{(s^{2} - 35.134s + 10639.2886)(s^{2} - 84.766s + 3571.2569)}$}
    \end{equation}
    
    \item \textbf{Circuit Components:} Comparing equation $(14)$, with the equation $(7)$ for the Second-order filter components, we get, \textbf{1st Second-Order Filter:} $R=10k\Omega$, $C_{1}=569.29nF$, $C_{2}=47.774nF$, \textbf{2nd Second-Order Filter:} $R=10k\Omega$, $C_{1}=235.95nF$, $C_{2}=118.68nF$

\end{itemize}

\subsection{Circuit Design}
The computed components were assembled to create a Fourth-order Chebyshev-I analog low-pass filter. The following circuit was constructed
with components, and was tested in the Hardware Lab and also with
software simulations.

\begin{figure}[H]
    \centering
    \includegraphics[width=.8\columnwidth]{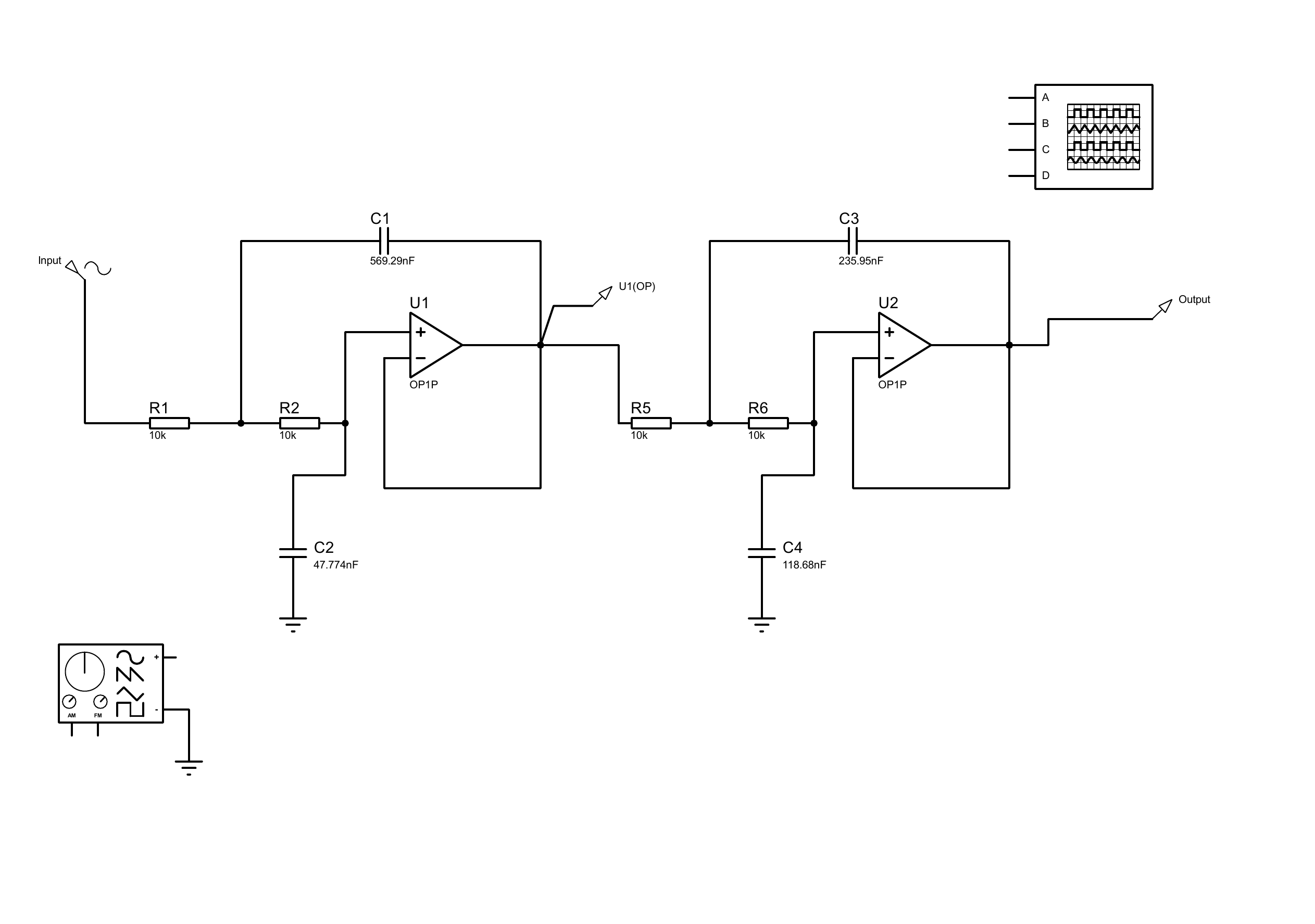}
    \caption{Analog Low Pass Chebyshev-I Filter designed with the computed components}
    \label{Chebyshev-I Filter Circuit}
\end{figure}

\subsection{Software Implementation}
\subsubsection{Proteus Implementation}

\begin{figure}[H]
    \centering
    \subfloat[\centering Circuit Simulation in Proteus 8]{{ \includegraphics[width=.5\columnwidth]{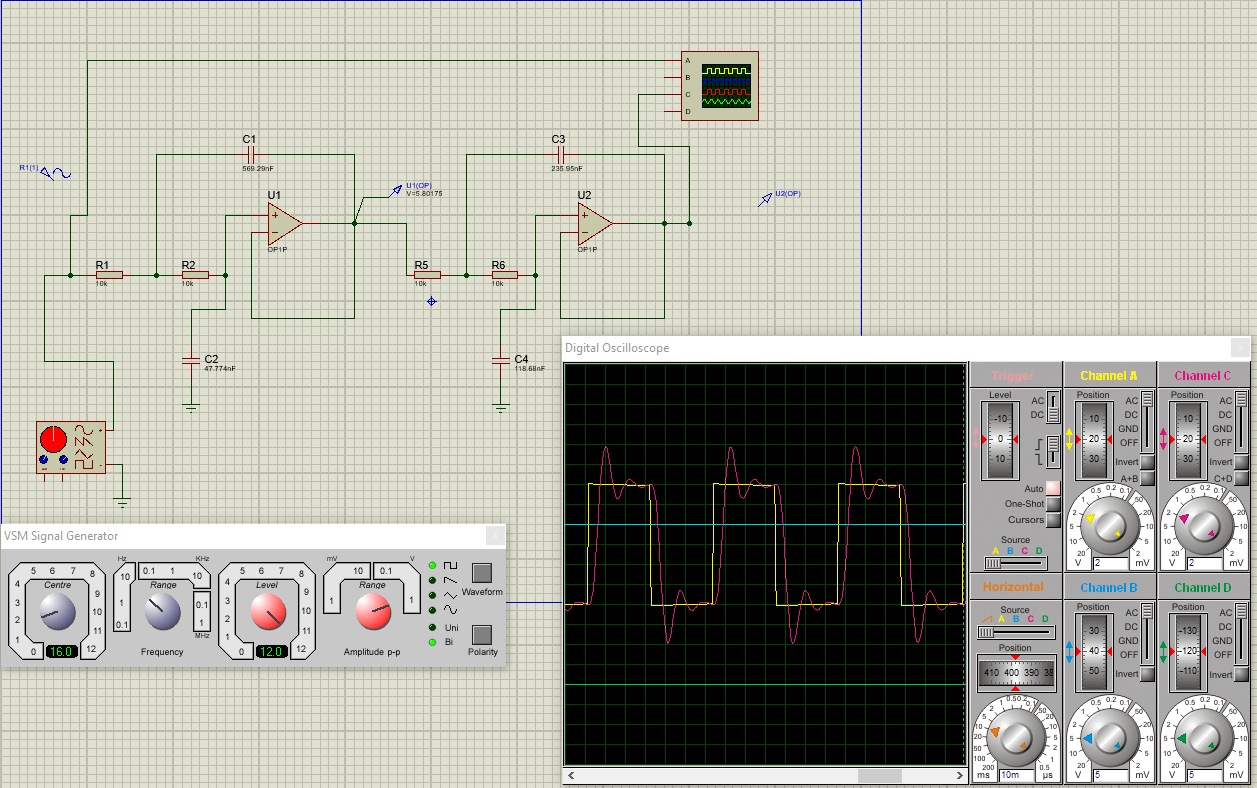} }}%
    \qquad
    \subfloat[\centering Output at the Output Voltage End in Proteus 8]{{ \includegraphics[width=.3\columnwidth]{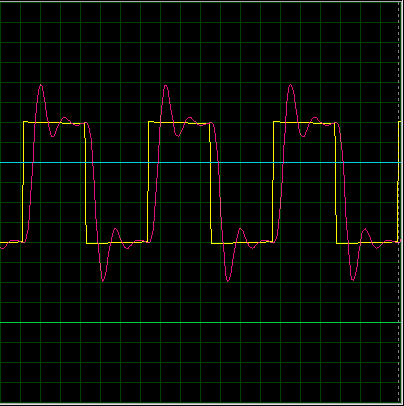} }}%
    \qquad
    \subfloat[\centering Simulated Frequency Response of the Designed Chebyshev-I Filter]{{\includegraphics[width=.5\columnwidth]{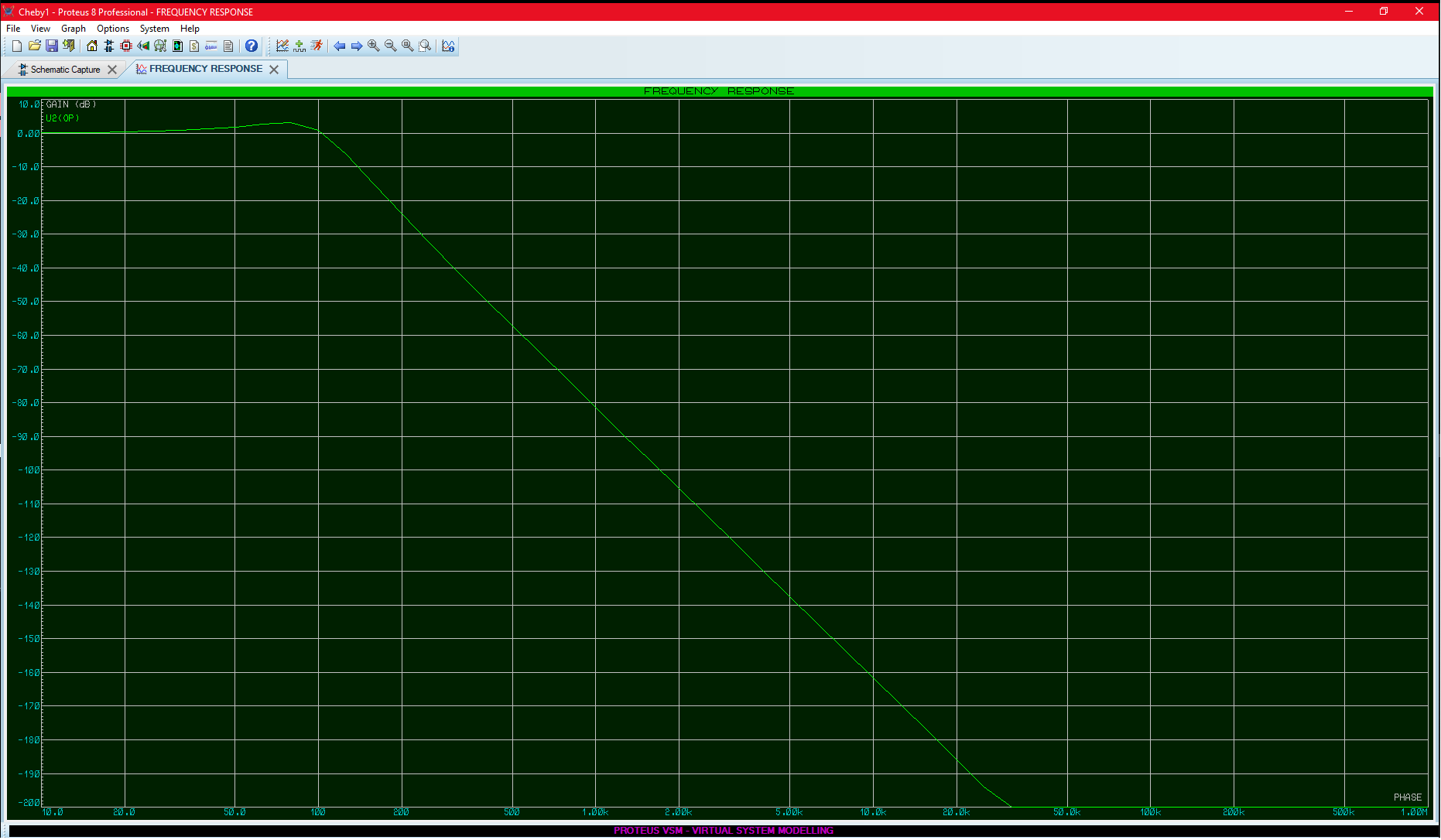} }}%
    \caption{Software Outputs}%
    \label{Chebyshev}%
\end{figure}

The circuit is constructed in the proteus environment, using the mathematically computed filter components. The circuit is then connected to the input voltage signal generator, where a square signal is fed. The value of input frequency was 15.91Hz $\approx$ 100 rad/s. The circuit's output voltage may be observed using a digital oscilloscope, which gives the desired result. The Frequency Response shows desired output by the simulation as well.

We can conclude by saying that our circuit simulation produced the desired output.

\subsubsection{Python}
\begin{itemize}
    \item \textbf{Order and Cut-off Frequency Determination:}
    
    \textbf{scipy.signal.cheb1ord} function has been used to determine the order of the filter $n$. Using this function we get, \textbf{Order of the Filter}, $N:\:4$. The results match the mathematically calculated results. 
    
    \item \textbf{Pole-Zero Location Determination:} 
    
    \textbf{scipy.signal.cheby1} function has been used to get the pole, zero locations of the Chebyshev-I filter. Using this we get, \textbf{Pole Locations}: $-19.42936433 \pm 103.61354304j$, $-46.90663488 \pm 42.91813477j$.
    
    \item \textbf{Pole and Zero Location Mapping:} \textbf{plotPoleZeros} function is created to map the pole-zero location on an S-plane.
    
    The results show the stable pole locations on the left of the imaginary axis, which matches our calculations distinctively. As the order of the filter is even, there will be no pole on the $x$-axis. So, the mapping is correct, as there are no poles on the $x$-axis.
    
    \item \textbf{Frequency Response Mapping:} \textbf{scipy.signal.freqs\_zpk} function has been used to get the frequency response of the Chebyshev-I filter designed, which shows accurate results.
    
    \begin{figure}[H]
    \centering
    \subfloat[\centering Pole Zero Location mapping with \textbf{plotPoleZeros} function]{{ \includegraphics[width=.4\columnwidth]{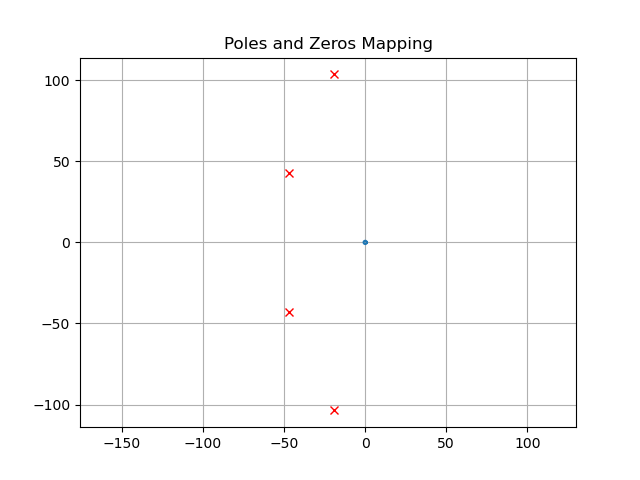} }}%
    \qquad
    \subfloat[\centering Frequency Response of the Chebyshev-I Filter]{{\includegraphics[width=.4\columnwidth]{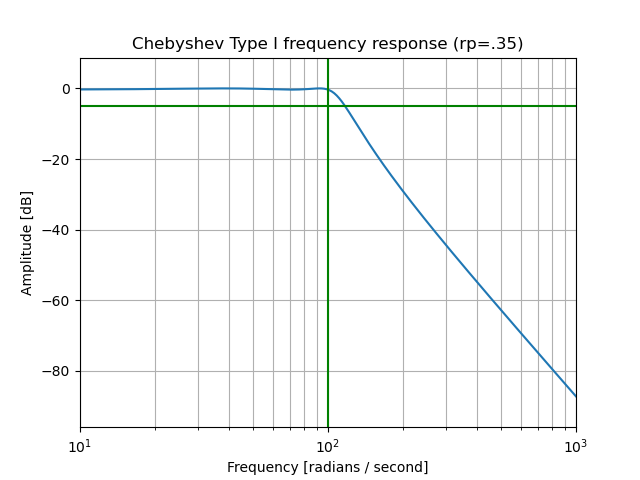} }}%
    \caption{Python Outputs}%
    \label{Chebyshev}%
    \end{figure}
    
\end{itemize}

\subsection{Hardware Implementation}
The developed circuit was implemented with hardware components in the lab, which provided the following voltage output values.

\begin{figure}[H]
    \centering
    \subfloat[\centering Output at input frequency of 5 Hz]{{ \includegraphics[width=.4\columnwidth]{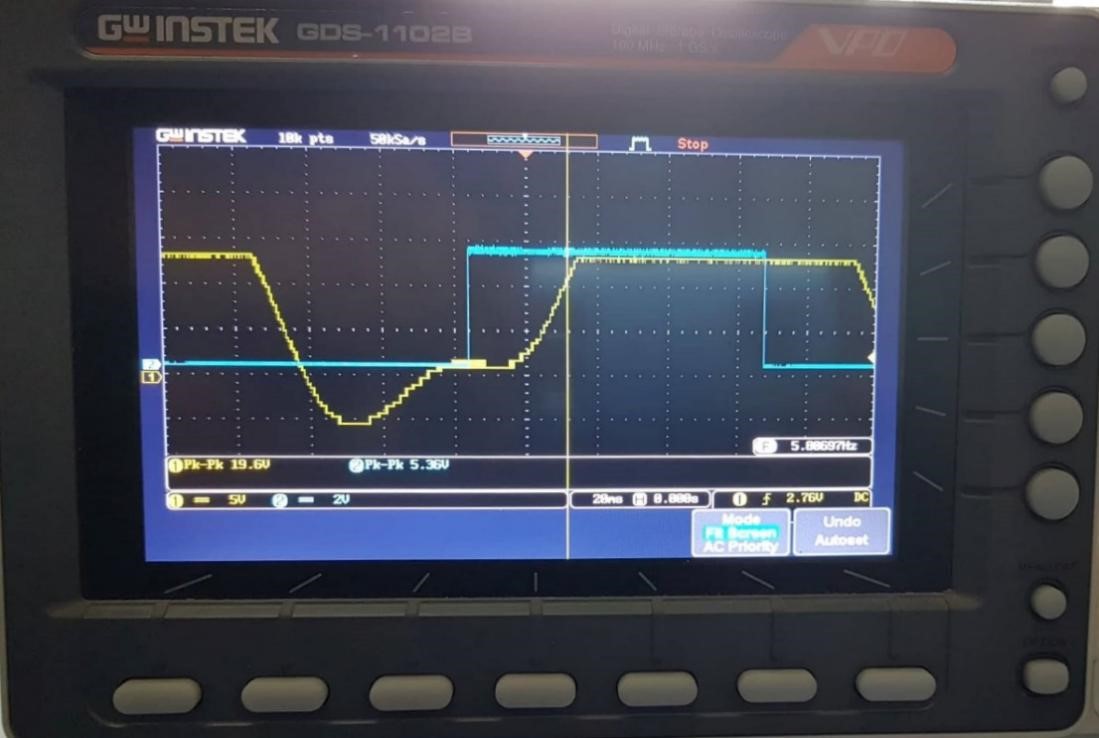} }}%
    \qquad
    \subfloat[\centering Output at input frequency of 13 Hz]{{\includegraphics[width=.4\columnwidth]{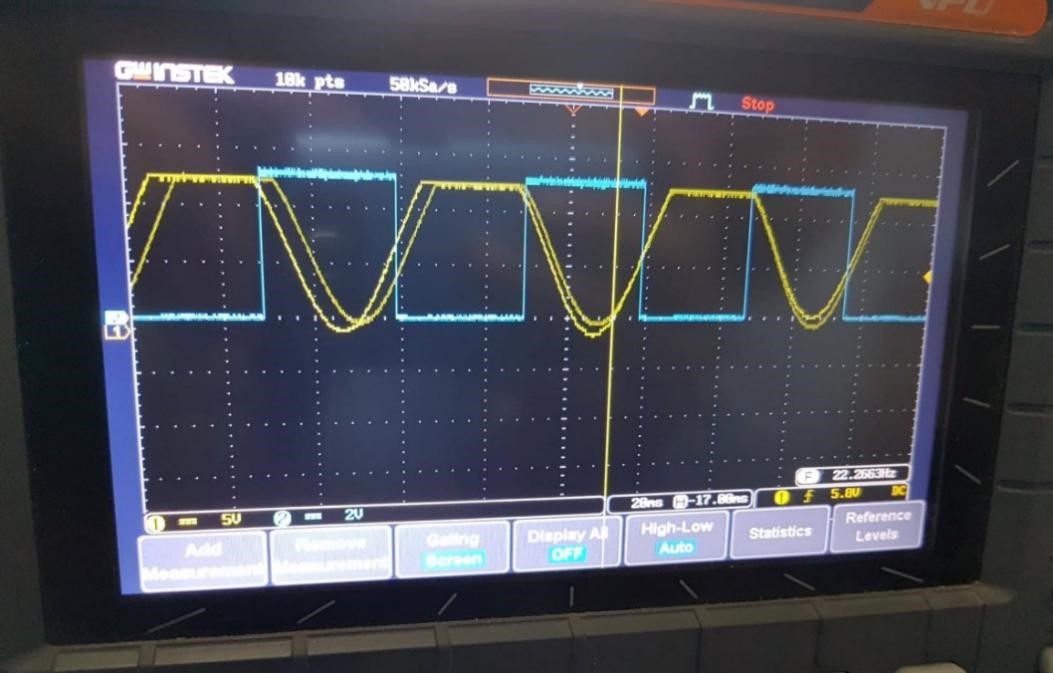} }}%
    \qquad
    \subfloat[\centering Output at input frequency of 20 Hz]{{\includegraphics[width=.4\columnwidth]{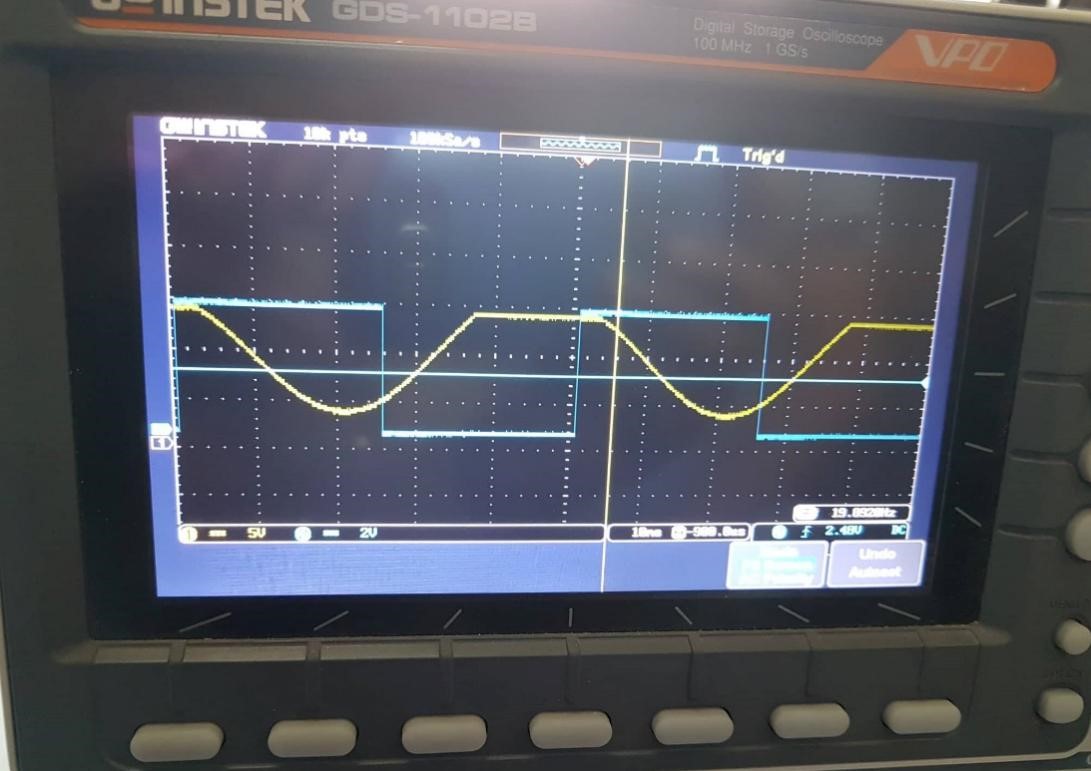} }}%
    \caption{Hardware Outputs (Blue = input, Yellow = output)}%
    \label{Butterworth}%
\end{figure}

\section{Comparison and Performance Analysis}
\subsection{Butterworth Filter Performance Analysis}
The Butterworth filter is the ideal solution between attenuation and phase reaction. It does not have a ripple in the passband or stopband, and because of this, it is often referred to as a maximum flat filter. The Butterworth filter achieves its flatness at the cost of a comparatively broad transition area from the passband to the stopband, with average transient characteristics.\cite{zumbahlen2011linear}\cite{administrator_2019}

\begin{itemize}
    \item A Butterworth filter does not have any ripple in the passband and the stopband. The designed filter does not have any ripples as well. 
    \item If we increase the order of the filter, it will move closer to its ideal response. But, if the order is increased, the amount of components increases as well.
    \item In the simulation a slight fluctuation in the passband can be seen, which is not a ripple or ringing. It is basically due to the approximation of the values of the capacitors. 
    \item We can see the sharp roll-off around the cut-off frequency, which is the desired outcome.
    \item Hardware implementation of the frequency response characteristics is slightly off due to the noise and other external occurrences, but gives similar results as the the software implementations.
\end{itemize}

Butterworth's ideal frequency response is unattainable as it creates an intense ripple of the passband.\cite{basicelectronicstutorials2018} Our Hardware and Software, both simulations provide similar results. We can conclude that the designed Butterworth filter gives the desired effect, both in Hardware and Software.

\subsection{Chebyshev-I Filter Performance Analysis}
The inclusion of the passband ripple as a parameter makes the Chebyshev filter's specification process a little more complicated than for the Butterworth filter, but it also increases versatility. The nominal gain of the filter is equal to the overall passband gain of the filter. Chebyshev Type I filters have ripple in the passband and monotone in the stopband. As such, the Type I filters roll off smoother than the Butterworth filters, but at the cost of a higher passband ripple.\cite{asn_home_2021}

\begin{itemize}
    \item Chebyshev-I filters have ripple in the passband. Our calculations show slight ripple in the passband, but this is due to the lower value of maximum ripple allowed $\varepsilon$. If we increase the value of $\varepsilon$, then we will find a large amount of ripple. But, as the calculation demands, the filter is correctly configured to give the expected response.
    \item As the characteristics of Chebyshev-I filter suggests, it provides maximally flat stopband.
    \item Comparatively has faster roll-off. But, it is at the cost of higher passband ripple. 
    \item The increase of the order of the filter will result in a more ideal response.
    \item Hardware implementation of the Frequency Response Characteristics is marginally off due to noise and other external occurrences, but achieves comparable effects as mathematical calculations and software implementations.
    \item Some Mathematical and Software calculations differ due to the approximation values, which can be overlooked. Esp. the pole-zero locations calculated manually and obtained from the python program differs slightly, which can be ignored. 
\end{itemize}

Both our hardware and software simulations produce very similar findings. However, we can infer that the designed Chebyshev-I filter gives the expected results, both in Hardware and Software.

\subsection{Comparison between Butterworth and Chebyshev-I Filter}
The Butterworth filters are designed to provide as flat a frequency response as possible in the passband.\cite{butterworth1930theory} On the other hand, Chebyshev-I filters are analog or digital filters with a roll-off steeper than Butterworth filters and have ripple in the passband. Through the experiment, the key comparison factors obtained are,

\begin{itemize}
    \item We can see that, for the same requirements, the number of poles in Butterworth $(n=5)$ is higher relative to the Chebyshev filter $(n=4)$, that is, the order of the Chebyshev filter is smaller than that of Butterworth. This is a tremendous benefit, since fewer discrete components would be used to create a filter.
    
    \item Chebyshev has a steeper roll-off near the cut-off frequency relative to Butterworth. While the monotony of the pass band is affected. This means that the Chebyshev filter at the same order attenuates the stop band frequencies in a better way than the Butterworth filter. Therefore, the same order Chebyshev low-pass filter will operate more efficiently than the Butterworth low-pass filter when disposing of undesirable frequencies, however, if the Butterworth filters, it could be a better alternative if a ripple-less and maximally flat solution is required. Thus, the application case of the filters determines the usability in certain situations.\cite{malica2011design}
    
    \item The Chebyshev filter has a larger decrease in magnitude with an increase in frequency than Butterworth. So, the derivative of Chebyshev's gain is going to be more detrimental than the Butterworth filter.
\end{itemize}

\begin{figure}[H]
    \centering
    \includegraphics[width=.6\columnwidth]{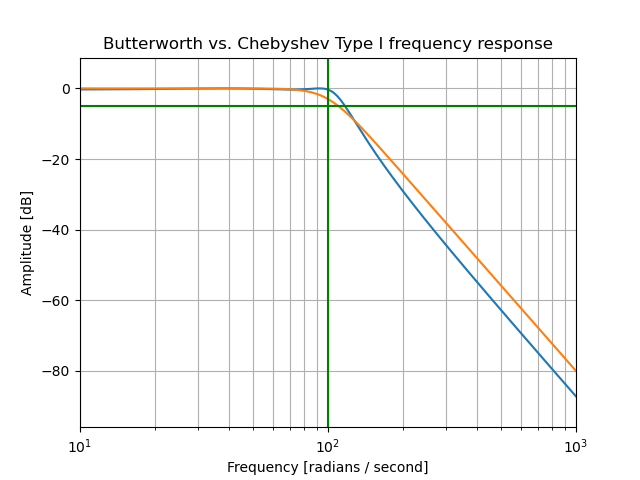}
    \caption{Comparison between the frequency response of Butterworth and Chebyshev-I Filters}
    \label{Frequency Response}
    \end{figure}

\section{Conclusion}
Butterworth and Chebyshev-I are two types of analog filters that are used for several purposes. This study's main objective is to design, implement, compare, and analyze these two filters correctly and efficiently. The results obtained from the Mathematical Calculations, Software and Hardware implementations are quite identical, suggesting that the tests have been carried out successfully and that the purpose of this research has been accomplished.

%

%


%
\section*{Conflict of Interest}

The authors declare that they have no conflict of interest.

\section*{Code Availability}

Available upon request.

\section*{Funding}

Not Applicable.

\section*{Availability of Data and Material}

Not Applicable.

\bibliographystyle{spmpsci}      
\bibliography{refs}   


\end{document}